\documentclass[12pt]{article}
\usepackage{amsmath,graphicx}
\usepackage[latin1]{inputenc}
\usepackage{hyperref}
\usepackage{xcolor}
\DeclareMathOperator\sinc{sinc}

\def\l({\left(}
\def\r){\right)}

\begin{document}

\title{{\bf New techniques to solve the 1-dimensional random flight}}
\author{
Ricardo Garc\'\i a-Pelayo\thanks{%
E-mail: r.garcia-pelayo@upm.es} \\
\\
ETS de Ingenier\'ia Aeron\'autica \\
Plaza del Cardenal Cisneros, 3 \\
Universidad Polit\'ecnica de Madrid \\
Madrid 28040, Spain\\}
\date{}
\maketitle

\begin{abstract}

We present here two new techniques to solve the one-dimensional random flight. The first one is an expansion in the number of collisions. The second one is the obtention of a Fourier series. This second technique can be applied to an arbitrary number of dimensions. Some mathematical identities are obtained.

\end{abstract}

\noindent{\bf{Keywords}}
\medskip

Telegrapher's equation; persistent random walk; Bessel functions; random flight

\tableofcontents

\centerline{}

\section{Introduction}

In the random flight a particle moves with a constant speed $v$ until, at times uniformly distributed with density $\lambda$ (that is, these times are a Poisson process with rate $\lambda$), it takes a uniformly distributed direction, while maintaining the same speed.

There are, of course, generalizations of this definition (see page xix of \cite{Kolesnik2021}). The random flight is the simplest case of ``continuous time random walk" (CTRW) (\cite{Montroll1965}, p. 177). The words ``random flight" may also be used when the changes of direction do not take place at uniformly distributed times but at equally spaced times. This usage is due to Rayleigh \cite{Rayleigh1919} (reprinted in \cite{Rayleigh1964}) as quoted in page 55 of \cite{Hughes1996}.

On the other hand, in the persistent random walk in 1 dimension, introduced by Fürth in 1920 \cite{Fuerth1920} and by Taylor in 1921 \cite{Taylor1921}, a particle moves with a constant velocity until, at each time step, it changes direction with probability $q$ and keeps the same direction with probability $p \equiv 1 - q$. In 1951 Goldstein considered the persistent random walk in which $p = 1 - (\lambda/2) dt$ and $q = (\lambda/2) dt$ as $dt \rightarrow 0$ \cite{Goldstein1951}. He found that the probability density function (henceforth pdf) of this persistent random obeyed the telegrapher's equation, first written by Heaviside in 1876 \cite{Heaviside1876} to address electricity problems raised by telegraphy transatlantic cables.

A little reflection \cite{RGPN1995} shows that the persistent random walk solved by Goldstein is equivalent to a random flight in which the collisions take place at times uniformly distributed with density $\lambda$.

The random flight in 1 dimension has been applied to Taylor dispersion \cite{Vandenbroeck1990}, option pricing \cite{Kolesnik2013}, surface gravity waves \cite{Caceres2021} and other diffusion problems \cite{Weiss2002}. We also bring to the reader's attention applications to relativistic Quantum Mechanics, which are not often mentioned \cite{Gaveau1984,Mugnai1992}.

Two new techniques to find the position pdf of the 1-dimensional random flight are presented in this article. The first one (section 2) is an expansion in the number of collisions. The second one is a solution by a Fourier series (section 3). The solution to the bullet initial condition, i. e., when the all particles move either to the right or to the left, is also discussed.

As a by-product, three mathematical identities are obtained. Matching expression \eqref{eq:rhoColl} (up to its singular part) with expression \eqref{dim1Goldstein}, and expression \eqref{eq:bullet2012} with expression \eqref{eq:bullet2022.1} yields two identities involving Bessel functions; expression \eqref{identity} is an identity which involves hypergeometric functions.

\section{Expansion of the 1-dim case in the number of collisions} \label{Expansion of the 1-dim case in the number of collisions}

The solution of the 1-dimensional random flight has already been given as an expansion in the number of collisions (see section 2 of \cite{MFranceschetti2007} or section 6.3 of \cite{RGPN2012cap}) obtained by a mathematical argument (section 2 of \cite{MFranceschetti2007}). In this section a simpler such expansion is derived by a direct argument.

In dimension 1, in an infinitesimal interval of time $dt$, we may assign a probability $1 - \lambda/2\ dt$ to keep the direction of motion and a probability $\lambda/2\ dt$ to change it (persistent random walk model), or, equivalently, we may assign a probability $1 - \lambda\ dt$ to remain unscattered and a probability $\lambda\ dt$ to be scattered isotropically \cite{RGPN1995} (diffusive model). In the diffusive model $\lambda$ is the average number of scatterings per unit time. Symbolically, in an interval of time $dt$,

\begin{equation}\label{eq:duality}
p({\rm{scattering}}) = \lambda\ dt\ \Leftrightarrow\ p({\text{direction reversal}}) = {\lambda \over 2}\ dt.
\end{equation}
We shall use this equivalence between the two models to our advantage. We choose the former model to find the pdf $\rho_r(x,t)$ at the point $x$ for a particle with the condition that it has reversed directions $r$ times at random times between 0 and $t$.

We have used a language typical of the partial differential equation approach to the 1-dimensional random flight to compare the above two models. In this approach, $dt$ is an infinitesimal, that is $\lambda dt \rightarrow 0$ for any $\lambda$. We now make some remarks about the accuracy of any mathematical solution to the random flight, whether based on a differential equation or not (as in both approaches presented here).  When the time $t$ considered is very short, then the number of collisions which take place may, relatively speaking, have a large relative error compared to $\lambda t$. If a relative error less than, say, 0.1 is wanted, then it must be $\lambda t \geq 100$, because the relative error is the standard deviation divided by the expected value, which is $1 \over \sqrt{c}$ in a Poisson process, where $c$ is the number of collisions. This is a rule of thumb, because the number of collisions is not the pdf, which is the quantity of interest in this article. Further discussion of this topic is beyond the scope of this article.

In order to derive $\rho_r(x,t)$, first we study uniformly distributed numbers in an interval of the real line.

\subsection{Distribution of random times in an interval} \label{Distribution of random points in an interval}

Consider $r$ uniformly distributed times between 0 and $t$ with chronological order

\begin{equation}\label{PermNat}
0 \leq t_1 \leq t_2 \leq ... \leq t_r \leq t.
\end{equation}
How is $t_j$ distributed? For fixed $t_j$ the Lebesgue measure of the set of times which satisfies inequality \eqref{PermNat} is

\begin{equation}\label{}
\int_0^{t_j}dt_{j-1} \int_0^{t_{j-1}} dt_{j-2} \cdots \int_0^{t_2} dt_1 \int_{t_j}^t dt_{j+1} \int_{t_{j+1}}^t dt_{j+2} \cdots \int_{t_{r-1}}^1 dt_r.
\end{equation}
It is straightforward to prove by induction that

\begin{equation}\label{simplex1}
\int_0^{t_j}dt_{j-1} \int_0^{t_{j-1}} dt_{j-2} \cdots \int_0^{t_2} dt_1 = {t_j^{j-1} \over (j-1)!}
\end{equation}
and, using this result,

\begin{equation}\label{simplex2}
\int_{t_j}^t dt_{j+1} \int_{t_{j+1}}^t dt_{j+2} \cdots \int_{t_{r-1}}^t dt_r =
\int_0^{t-t_j} du_{j+1} \cdots \int_0^{t-t_{r-1}} du_r = {(t-t_j)^{r-j} \over (r-j)!}.
\end{equation}
The multiplication of these results, \eqref{simplex1} and \eqref{simplex2}, yields ${(t-t_j)^{r-j} t_j^{j-1} \over (j-1)! (r-j)!}$. In order to find the pdf $\rho(r,t_j,t)$ of $t_j$ we still have to normalize. After normalization we obtain

\begin{equation}\label{pointdistr1}
\rho(r,t_j,t) = {r! \over (j-1)! (r-j)!}\ {(t-t_j)^{r-j} t_j^{j-1} \over t^r}.
\end{equation}

\subsection{Distribution of distance between random points in an interval} \label{Distribution of distance between random points in an interval}

What we have in mind is a random walker which moves with speed $v$ and which may, therefore, be at a distance at most $vt$ from the origin at tine $t$. In order to find the pdf of its position ar time $t$ we are going to use the pdf of the distance $d$ between two neighboring points when $r$ points have been tossed between 0 and $vt$. This pdf is

\begin{equation}\label{interpointdist2}
\rho(r,d) = r {(v t-d)^{r-1} \over (v t)^r}.
\end{equation}

Indeed, if $d$ is given and the two neighboring points are $x_j$ and $x_{j+1}$, $x_{j+1}$ is determined by $x_j$ and $d$. Thus $\rho(r,d)$ has to be proportional to the number of ways of accommodating the points different from $x_j$ in a length $vt-d$. Note that this argument also applies to the distance between 0 and $x_1$ or between $x_{r}$ and $vt$.


\subsection{Derivation of the pdf}

In the persistent random walk model with speed $v$, let the random times at which the $r$ reversals take place be $0<t_1<...<t_r<t$. Then the position at $t$ is $\pm v \big( (t_1-0) - (t_2-t_1) + ... +(-1)^{r-1} (t_r-t_{r-1}) +(-1)^{r} (t-t_r) \big)$, the sign before $v$ depending on whether the first move was to the right or to the left. The terms $(t_{i+1}-t_i)$ are identically distributed dependent random variables (in fact, distributed with the pdf \eqref{interpointdist2}). Therefore we can freely interchange the signs of the terms as long as the total number of $-$ and + signs is kept constant. Then the position is distributed as $x \equiv \pm v \Big( \big(t_1-0\big) + \big(t_2-t_1\big) + ... + \big(t_{\lfloor{r+1 \over 2}\rfloor}-t_{\lfloor{r-1 \over 2}\rfloor}\big) - \big(t_{\lfloor{r+3 \over 2}\rfloor}-t_{\lfloor{r+1 \over 2}\rfloor}\big) -...-\big(t-t_{r}\big) \Big) = \pm v \Big( t_{\lfloor{r+1 \over 2}\rfloor} - \big(t-t_{\lfloor{r+1 \over 2}\rfloor}\big) \Big) = \pm v \Big( 2 t_{\lfloor{r+1 \over 2}\rfloor} - t \Big)$. Its distribution, which we denote by $\rho_r(x,t)$, can be obtained from formula \eqref{pointdistr1} by the change of variable

\begin{equation}\label{}
x = \pm v \Big( 2 t_{\lfloor{r+1 \over 2}\rfloor} - t \Big)\ \Leftrightarrow\ t_{\lfloor{r+1 \over 2}\rfloor} = {1 \over 2} \Big( t \mp {x \over v} \Big)
\end{equation}
and the relation

\begin{equation}\label{}
j = \left\lfloor{r+1 \over 2}\right\rfloor.
\end{equation}
We introduce the Heaviside function $H$, which is equal to its argument when the argument is positive, and 0 otherwise. We are to find the solution of the isotropic case, for which the two initial signs are equally likely:

$$
\rho_r(x,t) =
$$

$$
\Bigg(
{1 \over 2} \left| {d t_{\lfloor{r+1 \over 2}\rfloor} \over dx } \right| \rho\Big(r, {1 \over 2} \Big( t - {x \over v} \Big), t \Big) + {1 \over 2} \left| {d t_{\lfloor{r+1 \over 2}\rfloor} \over dx } \right| \rho\Big(r, {1 \over 2} \Big( t + {x \over v} \Big), t \Big) \Bigg) H(v t - |x|) =
$$

$$
\Bigg(
{1 \over 2} \left| - {1 \over 2v } \right| {r! \over \lfloor{r-1 \over 2}\rfloor! \big( r - \lfloor{r+1 \over 2}\rfloor \big)!}\ {\left( {1 \over 2} \Big( t - {x \over v} \Big) \right)^{\lfloor{r-1 \over 2}\rfloor} \left( {1 \over 2} \Big( t + {x \over v} \Big) \right)^{r - \lfloor{r+1 \over 2}\rfloor} \over t^r} +
$$

$$
{1 \over 2} \left| - {1 \over 2v } \right| {r! \over \lfloor{r-1 \over 2}\rfloor! \big( r - \lfloor{r+1 \over 2}\rfloor \big)!}\ {\left( {1 \over 2} \Big( t + {x \over v} \Big) \right)^{\lfloor{r-1 \over 2}\rfloor} \left( {1 \over 2} \Big( t - {x \over v} \Big) \right)^{r - \lfloor{r+1 \over 2}\rfloor} \over t^r}  \Bigg) H(v t - |x|) =
$$

$$
{1 \over 4 v t^r} \left\lfloor {r+1 \over 2} \right\rfloor {r \choose \lfloor {r+1 \over 2} \rfloor}
\left( {v^2 t^2 - x^2 \over 4 v^2} \right)^{\lfloor{r-1 \over 2}\rfloor}
$$

$$
\Bigg( \left( {1 \over 2} \Big( {vt + x \over v} \Big) \right)^{r - \lfloor{r+1 \over 2}\rfloor - \lfloor{r-1 \over 2}\rfloor} +
       \left( {1 \over 2} \Big( {vt - x \over v} \Big) \right)^{r - \lfloor{r+1 \over 2}\rfloor - \lfloor{r-1 \over 2}\rfloor} \Bigg) H(v t - |x|) =
$$

$$
{1 \over 4 v} \left\lfloor {r+1 \over 2} \right\rfloor {r \choose \lfloor {r+1 \over 2} \rfloor}
\left( {v^2 t^2 - x^2 \over v^2 t^2} \right)^{\lfloor{r-1 \over 2}\rfloor} {1 \over t^{r-2 \lfloor{r-1 \over 2}\rfloor} 2^{2 \lfloor{r-1 \over 2}\rfloor}}
$$

$$
\Bigg( \left( {1 \over 2} \Big( {vt + x \over v} \Big) \right)^{r - 1 - 2 \lfloor{r-1 \over 2}\rfloor} +
       \left( {1 \over 2} \Big( {vt - x \over v} \Big) \right)^{r - 1 - 2 \lfloor{r-1 \over 2}\rfloor} \Bigg) H(v t - |x|) =
$$

%

\begin{equation}\label{}
{1 \over 2^r v t} \left\lfloor {r+1 \over 2} \right\rfloor {r \choose \lfloor {r+1 \over 2} \rfloor} \left({  v^2 t^2 - x^2 \over (v t)^2  } \right)^{\lfloor {r-1 \over 2} \rfloor} H(v t - |x|), \ \ \ \forall r=1,2,3,...
\end{equation}

Note that

\begin{equation}\label{Por parejas}
\forall r=1,2,3,...\ \ \ \rho_{2r-1}(x,t) = \rho_{2r}(x,t),
\end{equation}
that is, $\rho_{1}(x,t) = \rho_{2}(x,t) = {1 \over 2} {1 \over v t}$, $\rho_{3}(x,t) = \rho_{4}(x,t) = {3 \over 4} { (v t)^2 - x^2 \over (v t)^3}$,... This allows us to present the above result without using the floor notation.
\medskip

Since

\begin{equation}\label{}
\rho_{0}(x,t) = {1 \over 2} \big( \delta(x+v t) + \delta(x-vt) \big),
\end{equation}

and the number of direction reversals in the time interval $[0,t]$ is Poisson distributed with mean $1/\lambda$, it follows that

$$
\rho(t,x) = e^{-\lambda t/2} \sum_{r=0}^\infty {(\lambda t/2)^r \over r!} \rho_r(x,t) =
$$

$$
{e^{-\lambda t/2} \over 2} \big( \delta(x+v t) + \delta(x-vt) \big) +
$$

\begin{equation}\label{eq:rhoColl}
e^{-\lambda t/2} \sum_{n=1}^\infty {(\lambda t/2)^{2n-1} \over (n-1)!^2} \left( 1 + {\lambda t/2 \over 2n} \right) {1 \over 2^{2n-1} v t} \left( {v^2 t^2 - x^2 \over v^2 t^2} \right)^{n-1} H(v t - |x|).
\end{equation}
The $n$ index in the above expansion is not the number of collisions. The term of index $n$ corresponds to $2 n -1$ and $2 n$ direction reversals, because relations \eqref{Por parejas} have been used. The non-singular part of this expansion is equal to the solution $\rho_G$ obtained by Goldstein in 1951 \cite{Goldstein1951}:

$$
\rho_G(t,x) \equiv
$$

\begin{equation}\label{dim1Goldstein}
{\lambda e^{-\lambda t/2} \over 4 v} \left( I_0\bigg( {\lambda \sqrt{v^2 t^2 - x^2} \over 2 v} \bigg) + {v t \over \sqrt{v^2 t^2 - x^2} } I_1\left( \lambda {\sqrt{v^2 t^2 - x^2} \over 2 v} \right) \right) H(v t - |x|),
\end{equation}
where $I_0$ and $I_1$ are the Bessel functions of imaginary argument:

\begin{equation}
I_0(z) = \sum_{j=0}^\infty {({1 \over 2} z)^{2 j} \over (j!)^2},\ \
\  I_1(z) = \sum_{j=0}^\infty {({1 \over 2} z)^{2 j+1} \over j!\
(j+1)!}.
\end{equation}
The pdf of the particles which have never changed direction is

\begin{equation}\label{NeverBack}
\eta_{unreversed}(x, t) = {e^{-\lambda t/2} \over 2} \big( \delta(x+vt) + \delta(x-vt) \big).
\end{equation}
Then

\begin{equation}\label{}
\rho(x,t) - \eta_{unreversed}(x, t) = \rho_G(x,t).
\end{equation}

\subsubsection{Bullet initial condition}

As written in the Introduction, the problem at hand can be cast as an equation in partial derivatives, which was solved by Goldstein \cite{Goldstein1951}. A method to solve this equation for arbitrary initial conditions using Green's functions is given in section 7.4 of reference \cite{Morse1953}, but the analytic form of the solution is not given.

So far we have found the pdf with the initial condition that the first step is taken with equal probability to the right or to the left. Now we find the pdf with the initial condition that the first step is taken to the right (or to the left). Any initial condition can be put in terms of the latter case.

As in reference \cite{RGPN2012cap} we denote by ``bullet initial condition" the initial conditions

\begin{equation}
\rho_{b\pm}(x,0) = \delta(x)\ \ \ {\rm{and}}\ \ \ {\partial \rho_{b\pm} \over \partial t} (x,0) =
\left.{\partial \delta (x \mp v t) \over \partial t}\right|_{t=0} = \mp v \delta' (x).
\end{equation}
In the case $\rho_{b+}(x,0)$, its pdf is $\delta(x - v t)$ until it scatters at time $t=0$. It follows straightforwardly from the previous discussion that in that case the pdf $\rho_{b+}$ can be obtained from

$$
\rho_{b+,c}(x,t) \equiv
$$

\begin{equation}\label{}
{1 \over 2 (v t)^c} {c! \over \big\lfloor{c-1 \over 2}\big\rfloor! \big\lfloor{c \over 2}\big\rfloor!}\ \Big({ v t - x \over 2} \Big)^{\lfloor{c \over 2}\rfloor} \Big({ v t + x \over 2} \Big)^{\lfloor{c-1 \over 2}\rfloor} H(v t - |x|), \ \ \ \forall c=1,2,3,...
\end{equation}
This expression must be substituted into the Poisson expansion. After some manipulations the following expression is obtained:

$$
\rho_{b+}(x,t) = e^{-\lambda t/2} \delta(x-vt) +
$$

\begin{equation}\label{eq:bullet2022.1}
e^{-\lambda t/2} \sum_{n=0}^\infty {(v t - x) \over 4 (v t)^{2n}}  {1 \over n!^2} \left( {\lambda t \over 2} \right)^{2 n} \left( {v^2 t^2 - x^2 \over v^2 t^2} \right)^{n-1} \left( n + {\lambda \over 4 v} (v t + x) \right) H(v t - |x|).
\end{equation}
This expression matches the solution (formula (79) of reference \cite{RGPN2012cap}):

$$
\rho_{b+}(x,t) = e^{ -{\lambda t/2}}\ \delta (x - v t) + {1 \over 2}\ e^{-{\lambda t/2}}
$$

\begin{equation}\label{eq:bullet2012}
\Bigg[{1 \over v} {{\lambda \over 2}}\ I_0\left({\lambda \over 2} \sqrt{t^2 - {x^2 \over v^2}} \right) +
\left({\lambda \over 2}\right)^2 {v t + x \over v^2}\ {1 \over {\lambda \over 2} \sqrt{t^2 - {x^2 \over v^2}}} I_1\left({\lambda \over 2} \sqrt{t^2 - {x^2 \over v^2}}\right) \Bigg] H(vt-|x|)
\end{equation}
to the telegrapher's equation with initial conditions of $\rho_{b+}$.

For the left moving particle there are as expressions \eqref{eq:bullet2022.1} and \eqref{eq:bullet2012}, except that $x \rightarrow -x$.

\section{Solution by Fourier series}

A real, integrable function $f$ of support $[-vt, +vt]$ is not periodic, because it takes non-zero values over the support and is 0 elsewhere. Therefore it has a Fourier transform but not a Fourier series. We may, however, consider the function $fp$ which repeats itself periodically outside the said interval. To be formal,

\begin{equation}\label{PeriodizedFunction}
f_p(x) \equiv f\big( ((x+vt)\ {\rm{mod}}\ 2 vt)-vt,t \big)
\end{equation}
Note that on $[-vt,+vt]$, $f_p(x) = f(x)$. Then $fp$ has a Fourier series. The Fourier series has obvious computational advantages over the Fourier transform. It is intuitive that the coefficients of the Fourier series of $fp$ are going to be the values at the corresponding frequencies of the Fourier transform of $f$. Indeed,

\begin{equation}\label{}
{1 \over 2 vt} \int_{-vt}^{+vt}dx'\ fp(x') e^{- i 2 \pi h {x' \over vt}} =
{1 \over 2 vt} \int_{-vt}^{+vt}dx'\ f(x') e^{- i 2 \pi h {x' \over vt}},\ \ \ \forall h \in Z.
\end{equation}
When $fp$ is even and has period $2 v t$, this implies that

\begin{equation}\label{TFSF}
fp(x) =
{1 \over 2 v t} \l( \tilde f(0) + 2 \sum_{h=1}^{\infty} \tilde f \l( {h \over 2 v t} \r) \cos {2 \pi h \over 2 v t} x \r),
\end{equation}
where $\tilde f$ is the Fourier transform of $f$. This will be used in the Fourier series \eqref{dim1.1series} which follows.

We now choose the scattering model presented in section \ref{Expansion of the 1-dim case in the number of collisions}. The expanding 1-dimensional front of un-scattered particles is

\begin{equation}
\eta_s(x, t) = {e^{-\lambda t} \over 2} \big( \delta(x+vt) + \delta(x-vt) \big).
\end{equation}
Note that this is different from expression \eqref{NeverBack}, because at each scattering there is a probability 1/2 that the particles keep their direction. Thus, there are particles which have been scattered but have not reversed their direction.

The integral equation

\begin{equation}
\rho (x,t) = \eta_s (x,t) + \lambda \int_0^t dt' \int dx'\ \eta_s (x', t') \rho(x - x', t-t')
\end{equation}
is the statement that at a given $(x,t)$ a particle has either scattered or not, and that in the second case it scattered for the first time at some $(x',t')$ and at that place and time the process started again giving birth to $\rho$ with origin in $(x',t')$.

To undo the space convolution in the above equation we take the Fourier transform (denoted by $\tilde{}$ ) and to undo the time convolution we take the Laplace transform (denoted by $\hat{}$ ). The result is

\begin{equation}
\hat{\tilde{\rho}} = \hat{\tilde{\eta_s}} + \lambda \hat{\tilde{\eta_s}} \hat{\tilde{\rho}},
\end{equation}
from which $\hat{\tilde{\rho}}$ can be solved:

\begin{equation}\label{FourierLaplaceSolution}
\hat{\tilde{\rho}} = {\hat{\tilde{\eta_s}} \over {1-\lambda \hat{\tilde{\eta_s}}}}.
\end{equation}
The Fourier-Laplace inversion of $\hat{\tilde{\rho}}$, when possible, solves the problem. This approach was pioneered by Montroll and Weiss \cite{Montroll1965}.

%
The Fourier and Fourier-Laplace transforms of $\eta_s$ are

\begin{equation}\label{TFshell}
\tilde \eta_s(\nu, t) = {e^{-\lambda t} \over 2} \int dx\ e^{-i 2 \pi \nu x} \big( \delta(x+vt) + \delta(x-vt) \big) = e^{-\lambda t} \cos 2 \pi \nu v t.
\end{equation}
and

$$
{\hat{\tilde{\eta}}}_s (\nu, \omega) = \int_0^\infty dt\ e^{-(\lambda + \omega) t} \cos 2 \pi \nu v t =
{1 \over \omega + \lambda} \int_0^\infty du\ e^{-u} \cos {2 \pi \nu v u \over \omega + \lambda} =
$$

\begin{equation}\label{FourierLaplaceEta}
= {\omega + \lambda \over (\omega + \lambda)^2 + (2 \pi \nu v)^2}.
\end{equation}
Substitution of ${\hat{\tilde{\eta_s}}}(\nu, \omega)$ into the Fourier-Laplace transform of the solution \eqref{FourierLaplaceSolution} yields

$$
\hat{\tilde{\rho}} (\nu, \omega) =
{\omega + \lambda \over (\omega + \lambda) \omega + (2 \pi \nu v)^2 } =
{\omega + \lambda \over (\omega + \lambda/2)^2 + (2 \pi \nu v)^2 -(\lambda/2)^2} =
$$

\begin{equation}\label{}
{\omega + \lambda/2 \over (\omega + \lambda/2)^2 + (2 \pi \nu v)^2 -(\lambda/2)^2} +
{\lambda \over 2 \sqrt{(2 \pi \nu v)^2 -(\lambda/2)^2} } {\sqrt{(2 \pi \nu v)^2 -(\lambda/2)^2} \over (\omega + \lambda/2)^2 + (2 \pi \nu v)^2 -(\lambda/2)^2}.
\end{equation}
The inverse Laplace transform of the above can be found from tables of Laplace transform pairs (\cite{Gradshteyn1980}, p. 1144, 15. and 18.) and the so-called shift theorem (\cite{Gradshteyn1980}, p. 1143). It is

\begin{equation}\label{TF}
\tilde{\rho} (\nu, t) = e^{-(\lambda/2) t} \l( \cos \sqrt{(2 \pi \nu v)^2 -(\lambda/2)^2}\ t +
{\lambda \over 2 \sqrt{(2 \pi \nu v)^2 -(\lambda/2)^2} } \sin {\sqrt{(2 \pi \nu v)^2 -(\lambda/2)^2}}\ t \r).
\end{equation}
When $4 \pi \nu v/\lambda > 1$ the above expression remains real, because, $\forall x \in \Re$, $\cos i x = \cosh x$ and $\sin i x = i \sinh x$.

\begin{figure}[h]
\centering
  \includegraphics[width=0.5 \textwidth]{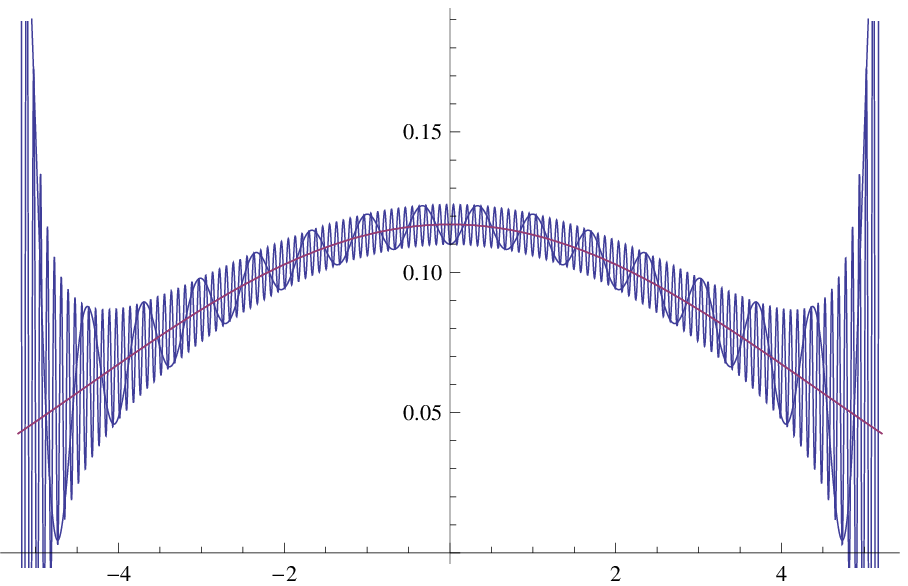}\includegraphics[width=0.5 \textwidth]{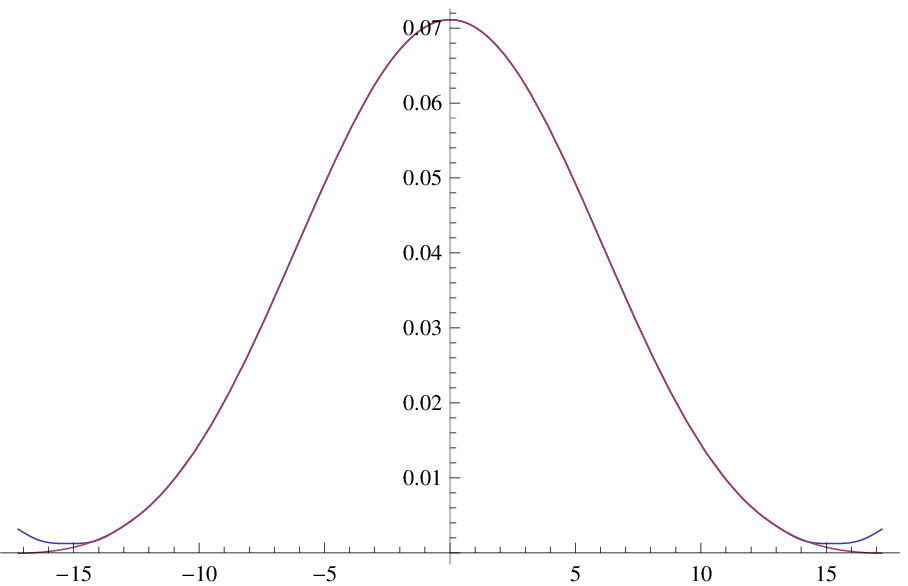}
  \caption{We set $\lambda = 1$ and $v = 1$, which is not a restriction on the cases, but just a choice of units of length and time. To the left, representation of the Fourier series \eqref{dim1.1series} with 15 and 125 terms. One can see that they have the same envelope. The non-wiggling arch in the middle of the oscillating graphs is $\rho_G$ \eqref{dim1Goldstein} for $t=5.21$ or the first 10 terms of $\rho_{cont}(x,5.21)$ \eqref{dim1.2series}. Their graphs overlay each other. To the right representation of the series \eqref{dim1.2series} with $\sum_{h=1}^{15}$ and of $\rho_G$ \eqref{dim1Goldstein} for $t=15.21$. }\label{banda}
\end{figure}

%
%
Then, according to relation \eqref{TFSF} and formula \eqref{TF},

$$
\rho(x,t) =
{1 \over 2 v t} \l( \tilde{\rho} (0, t) + 2 \sum_{h=1}^{\infty} \tilde{\rho} \bigg( {h \over 2 v t}, t \bigg) \cos {2 \pi h \over 2 v t} x \r) =
$$

%

\begin{equation}\label{dim1.1series}
{e^{-(\lambda/2) t} \over 2 v t} \Bigg( \cosh {\lambda t \over 2} + \sinh {\lambda t \over 2} +
2 \sum_{h=1}^{\infty} \l( \cos \sqrt{(\pi h)^2 -(\lambda t/2)^2} +
{\lambda t \over 2} \sinc {\sqrt{(\pi h)^2 - (\lambda t/2)^2}} \r) \cos {\pi h \over v t} x \Bigg).
\end{equation}
The Fourier series \eqref{dim1.1series} converges well when $t=15.21$. But as shown in Fig. \ref{banda}, not so when $t=5.21$, because oscillations appear about the solution $\rho_G$ \eqref{dim1Goldstein}. This is because the Fourier series of the expanding front \eqref{NeverBack} is oscillating. To get rid of these oscillations we simply remove the expanding front. Similarly to \eqref{TFshell},

\begin{equation}\label{TFshell2}
\tilde \eta_{unreversed}\bigg( {h \over 2 v t}, t \bigg) = e^{-\lambda t/2} \cos h \pi = (-1)^h e^{-\lambda t/2}.
\end{equation}
%
Thus, $\rho(x,t)$ without the expanding front is

$$
\rho_{cont}(x,t) =
{1 \over 2 v t} \l( \tilde{\rho}_{cont}  (0, t) + 2 \sum_{h=1}^{\infty} \tilde{\rho}_{cont} \bigg( {h \over 2 v t}, t \bigg) \cos {2 \pi h \over 2 v t} x \r) =
$$
%
%

$$
{e^{-(\lambda/2) t} \over 2 v t} \Bigg( \cosh {\lambda t \over 2} + \sinh {\lambda t \over 2} - 1+
$$

\begin{equation}\label{dim1.2series}
2 \sum_{h=1}^{\infty} \l( \cos \sqrt{(\pi h)^2 -(\lambda t/2)^2} +
{\lambda t \over 2} \sinc {\sqrt{(\pi h)^2 - (\lambda t/2)^2}} - (-1)^h \r) \cos {\pi h \over v t} x  \Bigg).
\end{equation}
Plots of $\rho_{cont}$ in $(-0.75 vt, +0.75 vt)$ with just 10 terms differ from those of $\rho_G$ by less than 1 part in 1,000 for all times. As $x$ approaches $\pm vt$, more terms are needed.



In regard to the convergence of the series \eqref{dim1.2series} for $\rho_{cont}$, we note that this series is a representation of $\rho_G$ \eqref{dim1Goldstein}, which is a continuous function of bounded continuous derivative. It follows then from the work of Dirichlet that its Fourier series converges pointwise (\cite{Koerner1989}, Theorem 1.1, p. 3).

We remark in Fig. \ref{banda} that the envelope of the oscillating series \eqref{dim1.1series} about the true solution will not diminish by including more frequencies. This shows that what we have here is not a case of the Gibbs' phenomenon. There is nothing particular about the time 5.21, it is just a time short enough for the oscillations to be quite visible, but not so short that they become huge and hide the fact that they roughly oscillate about the true value. The oscillations are still there for $t = 15.21$, but, as follows from formula \eqref{TFshell2}, their amplitude is $e^{-5} \approx 0.0067$ times smaller.

\subsection{Solution from the moments}

The moments of the pdf $\rho(x,t)$ are (formula (31) of \cite{RGPN2012cap}, which follows work done in \cite{RGPN1995})

\begin{equation}
\langle x^{2 m} \rangle =
{(2 m)! \over m!}\ e^{-{\lambda t}}\ (vt)^{2 m} \sum_{c=0}^\infty {(\lambda t)^c \over c!}\ {(m+c)! \over (2 m + c)!} = e^{-\lambda t}\ (vt)^{2 m} {}_1F_1(m + 1, 2 m + 1; \lambda t).
\end{equation}
The $m$-th moment is closely related to the $m$-th coefficient of the Maclaurin expansion of the Fourier transform:

\begin{equation}\label{}
\tilde \rho\bigg( {h \over 2 v t}, t \bigg)\equiv \int dx\ \rho(x,t)\ e^{-i {\pi h \over v t}x} =
\sum_{m=0}^\infty {(-i \pi)^m h^m \langle x^m\rangle(t) \over (v t)^m m!}.
\end{equation}
Since $\rho$ is even,

\begin{equation}\label{}
\tilde \rho\bigg( {h \over 2 v t}, t \bigg) = \sum_{m=0}^\infty {(-i \pi)^{2m} h^{2m} \langle x^{2m}\rangle(t) \over (v t)^{2m} (2m)!} =
e^{-\lambda t}\ \sum_{m=0}^\infty {(-i \pi)^{2m} h^{2m} \over (2m)!} {}_1F_1(m + 1, 2 m + 1; \lambda t).
\end{equation}
Therefore,

$$
\rho(x,t) = \sum_{h=-\infty}^{+\infty} {1 \over 2 v t} \tilde \rho\bigg( {h \over 2 v t}, t \bigg) \cos {2 \pi h \over 2 v t} x =
$$

\begin{equation}\label{dim1.3series}
{e^{-\lambda t} \over 2 v t} \sum_{h=-\infty}^{+\infty} \sum_{m=0}^\infty {(-i \pi)^{2m} h^{2m} \over (2m)!} {}_1F_1(m + 1, 2 m + 1; \lambda t) \cos {2 \pi h \over 2 v t} x.
\end{equation}
Comparison with expression \eqref{TF} and expansion \eqref{dim1.1series} yields the identity

$$
\sum_{m=0}^\infty {(-i \pi)^{2m} h^{2m} \over (2m)!} {}_1F_1(m + 1, 2 m + 1; \lambda t) =
$$

\begin{equation}\label{identity}
e^{(\lambda/2) t} \l( \cos \sqrt{(\pi h)^2 -(\lambda t/2)^2} +
{\lambda t \over 2} \sinc {\sqrt{(\pi h)^2 - (\lambda t/2)^2}} \r)
\end{equation}
From a numerical point of view the essential question is how many moments are needed to obtain a Fourier series coefficient with a prescribed error. That is, if $M$ is the upper limit in the sum over $m$, we want to know the function $M(h, \varepsilon_r)$, such that the truncation at $M$ yields the $h$-th Fourier series coefficient with a prescribed relative error $\varepsilon_r$. The function $M$ is, of course, a decreasing function of $\varepsilon_r$ and, not quite that obviously, an increasing function of $h$. The reason for the latter monotonous behavior is that in the identity \eqref{identity} the rhs, which is a function of $h$, is developed in powers of $h$ around 0 in the lhs. It is to be expected that the larger $h$ is, the more terms are needed in the lhs. A numerical exploration shows that for a relative error of $10^{-6}$, 69 moments suffice for $h \leq 15$ for all times.

As remarked before equation \eqref{TFshell2}, to get rid of the oscillations in expression \eqref{dim1.3series} due to the expanding front, the substitution

$$
e^{-\lambda t}\ \sum_{m=0}^\infty {(-i \pi)^{2m} h^{2m} \over (2m)!} {}_1F_1(m + 1, 2 m + 1; \lambda t) \rightarrow
$$

\begin{equation}\label{}
e^{-\lambda t}\ \sum_{m=0}^\infty {(-i \pi)^{2m} h^{2m} \over (2m)!} {}_1F_1(m + 1, 2 m + 1; \lambda t) - e^{-\lambda t/2} (-1)^h.
\end{equation}
must be made. Therefore,

$$
\rho_{cont}(x,t) =
{e^{-\lambda t} \over 2 v t} \bigg( e^{\lambda t} - e^{\lambda t/2}
$$

\begin{equation}\label{dim1.3series}
+ 2 \sum_{h=1}^{\infty} \bigg[ \bigg( \sum_{m=0}^\infty {(-1)^m (\pi h)^{2m} \over (2m)!} {}_1F_1(m + 1, 2 m + 1; \lambda t) \bigg) - (-1) e^{\lambda t/2} \bigg] \cos {\pi h \over v t} x \bigg).
\end{equation}

Note that when the Fourier series is obtained from the moments given in reference \cite{RGPN1995} nothing prevents us from applying the method in an arbitrary number of dimensions. Indeed, this has been done for the three-dimensional case \cite{RGPNtobe}.

%
%
%
%
%



\section{Conclusions}

The two expansions presented in this article have a clear a physical interpretation: an expansion in the number of collisions and an expansion in harmonics. We hope that they will be helpful and stimulating for the researchers working in dynamical processes.

We also hope that learning these two techniques will inspire further developments in other related problems, such as the ones mentioned in references \cite{Vandenbroeck1990}-\cite{Masoliver2019}. Two other related problems which we highlight are the hyperbolic diffusion in the presence of a force \cite{Caceres2021b} and the Boltzmann-Lorentz collision model \cite{Hauge1970,Caceres2022}. The bridging between the approach presented in this article and the Boltzmann-Lorentz collision model is promising but difficult for the following reason: when writing the Boltzmann equation for the Lorentz gas only the mean number of collisions appears \cite{Hauge1970,Caceres2022}, the distribution of the number of collisions is irrelevant.

\bigskip

{\Large{\bf{Acknowledgements}}}

This work was supported by MINECO/AEI and FEDER/EU under Project PID2020-112576GB-C21. The author thanks the MINECO/AEI of Spain for the financial support.


%

\end{document}